\documentclass[smallextended]{svjour3}       
\smartqed  
\usepackage{graphicx}
\usepackage{amsfonts,amssymb,amscd,amsmath,enumerate,verbatim,calc}
\usepackage[colorlinks,citecolor=blue]{hyperref}
\usepackage{color}

\spnewtheorem{result}{Result}{\bf}{\it}

\DeclareMathOperator{\grad}{grad}
\DeclareMathOperator{\tr}{trace}
\DeclareMathOperator*{\Ric}{Ric}
\DeclareMathOperator*{\diver}{div}

\newcommand{\co}{\nabla}
\newcommand{\D}{\partial}

\begin{document}

\title{Multiply-warped product metrics and reduction of Einstein equations}

\titlerunning{ }        

\author{F. Gholami,  F. Darabi,  A. Haji-Badali }

\authorrunning{F. Gholami \and F. Darabi    \and   A. Haji Badali          } 
\institute{F. Gholami \at
           Department of Mathematics, Faculty of Basic Sciences,   University of Bonab, Bonab, Iran\\
              \email{fateme.gholami@bonabu.ac.ir}
               \and
                F. Darabi (Corresponding author) \at
           Department of Physics, Azarbaijan Shahid Madani University, Tabriz, Iran\\
              \email{f.darabi@azaruniv.edu}
                   \and
        A. Haji-Badali  \at
              Department of Mathematics, Faculty of Basic Sciences,University of Bonab, Bonab, Iran\\
              \email{haji.badali@bonabu.ac.ir}
}

\date{Received: date / Accepted: date}

\maketitle

\begin{abstract}
It is shown that for every multidimensional metric in the multiply warped product form $\bar{M} = K\times_{f_1} M_1\times_{f_2}M_2$ with warp functions $f_1$, $f_2$,  associated to the submanifolds $M_1$, $M_2$ of dimensions $n_1$, $n_2$ respectively, one can find the corresponding Einstein equations $\bar{G}_{AB}=-\bar{\Lambda}\bar{g}_{AB}$, with cosmological constant $\bar{\Lambda}$, which are reducible to the Einstein equations $G_{\alpha\beta} = -\Lambda_1 g_{\alpha\beta}$ and $G_{ij} =-\Lambda_2 h_{ij}$ on the submanifolds $M_1$, $M_2$, with cosmological constants ${\Lambda_1}$ and ${\Lambda_2}$, respectively,
where $\bar{\Lambda}$, ${\Lambda_1}$ and ${\Lambda_2}$ are functions of ${f_1}$, ${f_2}$ and $n_1$, $n_2$.  
\end{abstract}

\section{Introduction}
Bishop and O'Niell  \cite{bishop.oneill:} were the pioneers who introduced the concept of warped products to construct examples of Riemannian manifolds
with negative curvature. Since then, the warped product spaces and their generic forms have been widely studied and used to construct new manifolds with interesting curvature properties. On the other hand, it was first noticed that in Lorentzian geometry some well known solutions to Einstein field equations, such as  Schwarzschild and Friedmann-Robertson-Walker metrics, can be expressed in terms of warped products. Therefore, Lorentzian warped products have also
been used to obtain more solutions to Einstein field equations.
Generalized Friedmann-Robertson-Walker spacetime and standard static
spacetime, as two well known solutions to Einstein field equations, can be also expressed as Lorentzian warped products. The former is a natural generalization of Friedmann-Robertson-Walker spacetime and the latter is a generalization of Einstein static universe. From mathematical point of view, the warped product geometries are useful when solving the partial differential equations over these backgrounds. This is because one can easily use and benefit of the method of separation of variables. From physical point of view, especially particle physics, the warped product geometries such as Randall-Sundrum models so called 5-dimensional warped geometry theory \cite{Randall.Sundrum:Analternative} are very important. In this theory, it is supposed that our real world is a higher-dimensional universe described by a warped geometry. More specifically, our universe is assumed to be a five-dimensional anti-de Sitter space where the elementary particles, except for the gravitons, are localized and confined
on the $(3 +1)$-dimensional brane embedded in the higher dimensional bulk spacetime.

In this paper, we consider a multiply warped product metric of the generalized Friedmann-Robertson-Walker spacetime type $\bar{M} = K\times_{f_1} M_1\times_{f_2}M_2$, with warp functions $f_1$, $f_2$ associated to the submanifolds $M_1$, $M_2$ with dimensions $n_1$, $n_2$ respectively. Then, we show that the Einstein equations $\bar{G}_{AB}=-\bar{\Lambda}\bar{g}_{AB}$ on the manifold $(\bar{M},\bar{g})$ with a cosmological constant $\bar{\Lambda}$ are reducible to the Einstein equations $G_{\alpha\beta} = -\Lambda_1 g_{\alpha\beta}$ and $G_{ij} =-\Lambda_2 h_{ij}$ on the submanifolds $(M_1, g)$ and $(M_2, h)$, with cosmological constants ${\Lambda_1}$ and ${\Lambda_2}$ respectively, such that $\bar{\Lambda}$, ${\Lambda_1}$ and ${\Lambda_2}$ are given in terms of the geometric features ${f_1}$, ${f_2}$ and $n_1$, $n_2$.
By using \cite{Bejancu:Classificationof5dwarpedspaceswithcosmologicalconstant,Gholami}, we consider some black hole solutions as typical examples and transform their metrics into the multiply warped product form of generalized Friedmann-Robertson-Walker metric $\bar{M} = K\times_{f_1} M_1\times_{f_2}M_2$ having warp functions $f_1$ and $f_2$. Then, we derive the corresponding Einstein equations $\bar{G}_{AB}=-\bar{\Lambda}\bar{g}_{AB}$, and the reduced Einstein equations $G_{\alpha\beta} = -\Lambda_1 g_{\alpha\beta}$ and $G_{ij} =-\Lambda_2 h_{ij}$ for each black hole solution.

\section{Preliminaries}
In this section, we state some definitions and results  of  \cite{Jaed,unal:curvature.multiply.warped} which will be needed throughout the current work.

Let $f$ be a smooth  function on a pseudo-Riemannian $n$-manifold $(M,g)$. Then  the Hessian tensor field of $f$ is given by
$H^f(X,Y)=XYf-(\co_XY)f,$
and
the Laplacian of $f$ is given by
$\Delta f=\tr_g(H^f),$ or equivalently, $\Delta = \diver(\grad),$  where $\co$, $\diver$ and $\grad$ are  the covariant derivative,  the divergence
and  the gradient oprators, respectively\cite[see p. 85 ]{oneil:book}.

Let $(B, g_B)$ and $(M_i, g_{M_i})$ be pseudo-Riemannian manifolds and
also let $n_i=\dim(M_i)$ and  $f_i : B\to(0,\infty)$ be smooth functions for any
$i \in \{1, 2, . . . , m\}$. The multiply-warped product is the
product manifold $ M = B \times M_1 \times M_2 \times ...\times M_m$
endowed with the metric tensor
$g = g_B \oplus f^2_ 1g_{M_1} \oplus f^2_2g_{M_2} \oplus ...\oplus
f^2_m g_{M_m}$ defined by
\begin{align} g = \pi^*(g_B) \oplus (f_1 o \pi)^2
\sigma^*_1(g_{M_1})\oplus...\oplus(f_m o \pi)^2\sigma^*_m(g_{M_m}),
\end{align}
where $\pi$ and $\sigma_i, (i=1, \ldots,m) $   are the natural projections of $B \times M_1 \times M_2 \times ...\times M_m$ onto $B$ and $M_1, \ldots,M_m$, respectively.

 Each function $f_i:B \to(0,\infty)$ is called a warping function and also each manifold $(M_i, g_{M_i} )$ is called a fiber manifold for
any $i \in \{1, 2, . . . , m\}$. The manifold $(B, g_B)$ is the base
manifold of the multiply-warped product.  The multiply-warped
product $(M, g)$ is a Lorentzian multiply-warped product if $(M_i,
g_{M_i})$ are all Riemannian for any $i\in\{1, 2,...,m\}$ and either
$(B, g_B)$ is Lorentzian or else $(B, g_B)$ is a one-dimensional
manifold with a negative definite metric $g_B=-dt^2$. If $B$ is an
open connected interval $I$ of the form $I = (t_1, t_2)$ equipped
with the negative definite metric $g_B =-dt^2$, where $t_1 < t_2$ ,
and $(M_i, g_{M_i})$ is Riemannian for any $i\in\{1, 2,...,m\}$,
then the Lorentzian multiply-warped product $(M, g)$ is called a
generalized Robertson-Walker spacetime or a multiply-warped
spacetime. In particular, a  generalized Robertson-Walker spacetime
is called a generalized Reissner-Nordstrom spacetime when $ m =
2$. {Since we have all information about the Levi-Civita connection and and the Riemann curvature tensor in \cite{unal:curvature.multiply.warped}, here we just recall from \cite{unal:curvature.multiply.warped} the Ricci curvature and the scalar curvature. }

\begin{pro}[\cite{unal:curvature.multiply.warped}]\label{1}
Let $M=B\times_{f_1}M_1\times\cdots\times_{f_m}M_m$ be a pseudo-Riemannian multiply-warped product with metric $g = g_B \oplus f^2_ 1g_{M_1} \oplus f^2_2g_{M_2} \oplus ...\oplus
f^2_m g_{M_m}$ also let $X,Y,Z\in \mathfrak{L}(B)$ and $V\in\mathfrak{L}(M_i), W\in\mathfrak{L}(M_j)$. Then
\begin{align}
\Ric(X,Y)&=\Ric^B(X,Y)-\sum_{i=1}^{m}\frac{n_i}{f_i}H_B^{f_i}(X,Y),\\
\Ric(V,X)&=0,\\
\Ric(V,W)&=0\mbox{ for } i\neq j,\\
\Ric(V,W)&=\Ric^{M_i}(V,W)-\bigg(\frac{\Delta_B f_i}{f_i}+(n_i-1)\frac{\mid\grad_B f_i\mid^2_B}{f_i^2}\nonumber\\
&\qquad+\sum_{k=1,k\neq i}n_k\frac{g_B(\grad_Bf_i,\grad_Bf_k)}{f_if_k}\bigg)g(V,W) \mbox{ for } i=j.
\end{align}
where $\Ric, \Ric^B, \Ric^{M_i}$ are the Ricci curvature   tensor  of metrics $g, g_B$ and $g_{M_i}$, respectively.
\end{pro}\\
\begin{pro}[\cite{unal:curvature.multiply.warped}]\label{2}
Let $M=B\times_{f_1}M_1\times\cdots\times_{f_m}M_m$ be a pseudo-Riemannian multiply-warped product with metric $g = g_B \oplus f^2_ 1g_{M_1} \oplus f^2_2g_{M_2} \oplus ...\oplus
f^2_m g_{M_m}.$ Then, scalar curvature $S$  of $(M,g)$ admits the following expressions
\begin{align}
S&=S^B-2\sum_{i=1}^mn_i\frac{\Delta_B f_i}{f_i}+\sum_{i=1}^m\frac{S^{M_i}}{f_i^2}-\sum_{i=1}^mn_i(n_i-1)\frac{\mid\grad_B f_i\mid^2_B}{f_i^2}\nonumber\\
&\qquad-\sum_{i=1}^m\sum_{k=1,k\neq i}^mn_in_k\frac{g_B(\grad_Bf_i,\grad_Bf_k)}{f_if_k}.
\end{align}
where $S^B$ and $ S^{M_i}$ are the scalar curvature  of metrics $ g_B$ and $g_{M_i}$, respectively.
\end{pro}
\section{ Generalized Friedmann-Robertson-Walker spacetime}
The Friedmann-Robertson-Walker (FRW) metric is an exact solution of Einstein's field equations, in four dimensional spacetime, which describes a homogeneous, isotropic expanding or contracting universe that may be simply connected or multiply connected. The general form of this metric follows from the geometric properties of homogeneity and isotropy of spacetime and is given by
\begin{equation}
\bar{g}(x^a)=\varepsilon dt^2 +f^2(t)g_{\alpha\beta}(x)dx^\alpha dx^\beta,
\end{equation}
where $\alpha,\beta,...\in  \{1,..., 3\}$. The 3-dimensional space has uniform curvature and can be an elliptical, Euclidean, or hyperbolic space. Einstein's field equations are only needed to derive the scale factor $f(t)$ of the universe as a function of cosmic time $t$. Although it is originally
obtained as a four dimensional metric, however, one can easily generalize this metric to higher dimensional spacetimes. This is well motivated by
high energy physics, especially string and brane theories, where it is supposed that the real numbers of spacetime dimensions may be higher than four, and the extra dimensions may be compactified or described in the
framework of warped product metrics \cite{Randall.Sundrum:Analternative}.

\begin{defi}
Let $(M_1,g_1)$ and $(M_2,g_2)$ be two Riemannian manifolds and $K$ a
1-dimensional manifold. Also let $f_i:K\rightarrow(0,\infty)$ be
smooth functions for any $i\in \{1,2\}$. The Lorentzian multiply-warped product is the product manifold $\bar{M} = K\times M_1\times
M_2$ furnished with the metric tensor $\bar{g}$ on $\bar{M}$ defined
by
\begin{align}
\bar{g}(x^a)=\varepsilon dt^2 +f^2_1(t)g_{\alpha\beta}(x^\mu)dx^\alpha dx^\beta +f^2_2(t)g_{ij}(x^k)dx^idx^j,
\end{align}
with local components
\begin{equation}\label{eq:metric1}
\left\{
\begin{array}{ll} \bar{g}_{00}=\bar{g}(\partial_t,\partial_t)=\varepsilon,
\\
\bar{g}_{\alpha\beta}=(f_1(t))^2{g_{1}}_{\alpha\beta}(x^\mu),
\\
\bar{g}_{ij}=(f_2(t))^2{g_2}_{ij}(x^k),
\\
\bar{g}_{i\alpha}=0,\  \bar{g}_{0i}=0,
\end{array}\right.
\end{equation}
where  $\varepsilon=\pm1$, \ $(x^\mu)$,\        $(x^k)$ and $t$ are
coordinate systems on $ M_1$, $M_2$ and $K$, respectively. We use
the ranges for the indices as $\alpha,\beta,...\in  \{1,..., n_1\},
i, j, ... \in \{n_1+1,...n_1+n_2\}$ and $a,b,c,...\in
\{1,...,n_1+n_2\}$ however we used $\D_t=\frac{\D}{\D t},
\D_i=\frac{\D}{\D x^i}$ and $\D_\alpha=\frac{\D}{\D x^\alpha}$. We
also define $f'_1=d f_1/dt $, $f'_2=d f_2/dt $, $ B_1=\frac{2f'_1}{f_1}, B_2=\frac{2f'_2}{f_2}$.

 \end{defi}
{ Now we can immediately  conclude the Ricci tensor and the scalar curvature of the generalized Friedmann-Robertson-Walker spacetime by a straightforward computation according to  \cite{unal:curvature.multiply.warped} and essentially the proposition \ref{1} and \ref{2}. }
\begin{pro}
Let $\bar{M}=K\times_{f_1} M_1\times_{f_2} M_2$ be a Lorentzian
multiply-warped products of $(M_1,g_1)$ and $(M_2,g_2)$.    Then we
have\begin{align}
\bar{\Ric}(\partial_t,\partial_t)&=-\left(m(\frac{B^2_1}{4}+\frac{B'_1
}{2})
+n(\frac{B^2_2}{4}+\frac{B'_2 }{2})\right),\label{eq:ric1}\\
\bar{\Ric}(\partial_\alpha,\partial_\beta)&=
\Ric^{M_1}(\partial_\alpha,\partial_\beta)-\bar{g}_{\alpha\beta}\left(\varepsilon(\frac{B^2_1}{4}+\frac{B'_1
}{2})
+(m-1)\varepsilon\frac{B^2_1}{4}+m\varepsilon\frac{B_1B_2}{4}\right),\label{eq:Ric2}\\
\bar{\Ric}(\partial_i,\partial_j)&=\Ric^{M_2}(\partial_i,\partial_j)-\bar{g}_{ij
}\left(\varepsilon(\frac{B^2_2}{4}+\frac{B'_2 }{2})+(n-1)\varepsilon\frac{B^2_2}{4}+n\varepsilon\frac{B_1B_2}{4}\right),\label{eq:ric3}\\
\bar{\Ric}(\partial_t,\partial_\alpha)&=0,\\
\bar{\Ric}(\partial_\alpha,\partial_i)&=0,
\end{align}
where $\Ric^{M_i}(\partial_\alpha,\partial_\beta)$ is the
local components of Ricci tensor of $(M_i,g_i)$.
\end{pro}
\begin{pro}
Let $\bar{M}=K\times_{f_1} M_1\times_{f_2} M_2$ be a Lorentzian multiply-warped products of $(M_1,g_1)$ and $(M_2,g_2)$.  Then, the
scalar curvature $\bar{S}$ of $(\bar{M},\bar{g})$ admits the following
expression
\begin{align}\begin{split}
\bar{S}= -2\left(m(\frac{B^2_1}{4}+\frac{B'_1 }{2})+
n(\frac{B^2_2}{4}+\frac{B'_2
}{2})\right)+\frac{S^{M_1}}{f^2_1}+\frac{S^{M_2}}{f^2_2}\\-\left(m(m-1)\varepsilon\frac{B^2_1}{4}+n(n-1)
\varepsilon\frac{B^2_2}{4}\right)-mn\varepsilon
\frac{B_1B_2}{4},\end{split}\label{eq:Scalar}
 \end{align}
  where $S^{M_1}$ and  $S^{M_2}$ are the scalar curvatures of $(M_1,g_1)$  and $(M_2,g_2)$, respectively.
\end{pro}

{ Using the proposition 4, we have what is necessary for obtaining the Einstein gravitational tensor component. }
\begin{pro}
Let $\bar{G}$ be the Einstein gravitational tensor field of
$(\bar{M},\bar{g})$, then we have the following equations
\begin{align}
\bar{G}_{00}&=\frac{-1}{2}\left(\varepsilon\frac{S^{M_1}}{f^2_1}+\varepsilon\frac{S^{M_2}}{f^2_2}-m(m-1)
\frac{B^2_1}{4}-n(n-1)\frac{B^2_2}{4}-mn\frac{B_1B_2}{4}\label{eq:ein1}\right),\\
\bar{G}_{\alpha0}&=0,\bar{G}_{i0}=0,\bar{G}_{i\alpha}=0,\label{eq:ein1'} \\
\begin{split}\bar{G}_{\alpha\beta}&=G_{\alpha\beta}+\bar{g}_{\alpha\beta}\left((m-1)\varepsilon(\frac{B^2_1}{4}
+\frac{B'_1 }{2})(m-1)(\frac{m}{2}-1)\varepsilon\frac{B^2_1}{4}\right.\\
&\quad+m(\frac{n}{2}-1)\varepsilon\frac{B_1B_2}{4}
+n\varepsilon(\frac{B^2_1}{4}+\frac{B'_1 }{2})-\frac{S^M_2}{2f^2}
\left.+\frac{n(n-1)}{2}\varepsilon\frac{B^2_2}{4}\right),\end{split}\label{eq:ein2}\\
\begin{split}\bar{G}_{ij}&=G_{ij}+\bar{g}_{ij}\bigg(\varepsilon(n-1)(\frac{B^2_2}{4}+\frac{B'_2 }{2})+\varepsilon(n-1)
(\frac{n}{2}-1)\frac{B^2_2}{4}\\
&\quad+\varepsilon n(\frac{m}{2}-1)\frac{B_1B_2}{4}
+n\varepsilon(\frac{B^2_1}{4}+\frac{B'_1}{2})-\frac{S^M_1}{2f^2_1}
\left.+\frac{m(m-1)}{2}\varepsilon\frac{B^2_1}{4}\right),
\end{split}\label{eq:ein3}
\end{align}
where $G_{\alpha\beta}$ and $G_{ij}$ are the local components of the Einstein
gravitational tensor field $G$ of $(M_1, g_1)$ and
$(M_2, g_2)$, respectively.
\end{pro}
\begin{proof}
Using Einstein gravitational tensor field of ($\bar{M}$,$\bar{g}$), $\bar{G} = \bar{\Ric} - \frac{1}{2}\bar{S}\bar{g}$, we have
$\bar{G}_{00} = \bar{\Ric_{00}} - \frac{1}{2}\bar{S}\bar{g}_{00}$
and by using relations \eqref{eq:metric1}, \eqref{eq:ric1} and
\eqref{eq:Scalar}æ we obtain the Einstein equations \eqref{eq:ein1},
\eqref{eq:ein1'}, \eqref{eq:ein2} and \eqref{eq:ein3}.
\end{proof}
\begin{pro}
The Einstein equations on $(\bar{M},\bar{g)}$ with cosmological
term $\bar{\Lambda}$ are equivalent with the following
reduced Einstein equations
\begin{align}
\begin{split}\bar{\Lambda}&=\frac{-1}{2}\bigg(m(m-1+n)(\frac{B^2_1}{4}+\frac{B'_1 }{2})+n(n-1+m)(\frac{B^2_2}{4}+\frac{B'_2 }{2})\\
&\qquad-(m^2+n^2-mn)\frac{B_1B_2}{4}\bigg), \end{split}\label{Lambda} \\
G_{\alpha\beta}&=\bar{g}_{\alpha\beta}(\frac{m}{2}-1)\left((m-1)\frac{\acute{B}_1}{2}-m\frac{B_1B_2}{4}+n(\frac{B^2_2}{4}+\frac{B'_2 }
{2})\right), \label{G} \\
G_{ij}&=\bar{g}_{ij}(\frac{n}{2}-1)\left((n-1)\frac{\acute{B}_2}{2}-n\frac{B_1B_2}{4}+m(\frac{B^2_1}{4}+\frac{B'_1
}{2})\right).\label{G'}
\end{align}
\end{pro}
\begin{proof}
 Using \eqref{eq:ein1} and  Einstein equation
$\bar{G}=-\bar{\Lambda}\bar{g}$ we have
\begin{align}
\bar{\Lambda}=\frac{1}{2}\bigg(\frac{S^M_1}{f^2_1}+\frac{S^M_2}{f^2_2}-m(m-1)\frac{B^2_1}{4}-n(n-1)\frac{B^2_2}{4}
-mn\frac{B_1B_2}{4}\bigg).\label{eq:landa1}
\end{align}
By using \eqref{eq:ein2},  Einstein equation
$\bar{G}=-\bar{\Lambda}\bar{g}$, and \eqref{eq:landa1} we obtain
\begin{align}
G_{\alpha\beta}=-\bar{g}_{\alpha\beta}(\frac{S^M_1}{2f^2_1}-(m-1)\frac{B^2_1}{4}-m\frac{B_1B_2}{4}
+(m-1)(\frac{B^2_1}{4}+\frac{{B_1}'}{2})+n(\frac{B^2_2}{4}+\frac{{B_2}'}{2})).\label{eq:landa2}
\end{align}
By contracting \eqref{eq:landa2} with $g^{\alpha\beta}$ we obtain
\begin{align}
\frac{S^M_1}{f^2_1}=m(m-1)\frac{B^2_1}{4}+m^2\frac{B_1B_2}{4}-m(m-1)(\frac{B^2_1}{4}
+\frac{{B_1}'}{2})-mn(\frac{B^2_2}{4}+\frac{{B_2}'}{2}).\label{eq:landa3}
\end{align}
Also by using \eqref{eq:landa2} and \eqref{eq:landa3} we obtain
\begin{align}
G_{\alpha\beta}=\bar{g}_{\alpha\beta}(\frac{m}{2}-1)\left((m-1)\frac{{B_1}'}{2}-m\frac{B_1B_2}{4}+n(\frac{B^2_2}{4}+
\frac{{B_2}'}{2})\right).
\end{align}
Similarly, by using \eqref{eq:ein3},  Einstein equation
$\bar{G}=-\bar{\Lambda}\bar{g}$, and \eqref{eq:landa1} we obtain
\begin{align}
&\frac{S^M_2}{f^2_2}=n(n-1)\frac{B^2_2}{4}+n^2\frac{B_1B_2}{4}-n(n-1)(\frac{B^2_2}{4}
+\frac{{B_2}'}{2})-mn(\frac{B^2_1}{4}+\frac{{B_1}'}{2}),\label{eq:landa4}\\
\nonumber
&G_{ij}=\bar{g}_{ij}(\frac{n}{2}-1)\left((n-1)\frac{{B_2}'}{2}-n\frac{B_1B_2}{4}+m(\frac{B^2_1}{4}+\frac{{B_1}'}{2})\right).
\end{align}
Finally, by using \eqref{eq:landa1}, \eqref{eq:landa3} and
\eqref{eq:landa4} we obtain
\begin{align}\label{BarLambda}
\bar{\Lambda}&=\frac{-1}{2}(m(m-1+n)(\frac{B^2_1}{4}+\frac{B'_1
}{2})+n(n-1+m)(\frac{B^2_2}{4} +\frac{B'_2
}{2})\\ \nonumber
&-(m^2+n^2-mn)\frac{B_1B_2}{4}).
\end{align}
\end{proof}
\begin{pro}
The Einstein equations $\bar{G}=-\bar{\Lambda}\bar{g}$ on
$(\bar{M},\bar{g})$ with cosmological term $\bar{\Lambda}$
induce the Einstein equations $G_{\alpha\beta} = -\Lambda_1
g_{\alpha\beta}$ on $(M_1, g_{\alpha\beta})$ and $G_{ij} =
-\Lambda_2 h_{ij}$ on $(M_2, h)$.
\end{pro}
\begin{proof}
By using \eqref{eq:metric1}, \eqref{G} and \eqref{G'}, we obtain $G_{\alpha\beta} = -\Lambda_1
g_{\alpha\beta}$ on $(M_1, g_{\alpha\beta})$ and $G_{ij} =
-\Lambda_2 h_{ij}$ on $(M_2, h)$ where the cosmological terms
$\Lambda_1$  and $\Lambda_2$ are given by
\begin{align}
\Lambda_1=-f^2_1(\frac{m}{2}-1)\left((m-1)\frac{{B_1}'}{2}-m\frac{B_1B_2}{4}+n(\frac{B^2_2}{4}+\frac{{B_2}'}{2})\right),\label{44}\\
\Lambda_2=-f^2_2(\frac{n}{2}-1)\left((n-1)\frac{{B_2}'}{2}-n\frac{B_1B_2}{4}+m(\frac{B^2_1}{4}+\frac{{B_1}'}{2})\right).\label{45}
\end{align}
\end{proof}

Requiring $\bar{\Lambda}$, ${\Lambda}_1$, and ${\Lambda}_2$ to be constant
(cosmological constant), leads to a set of coupled differential equations (\ref{BarLambda}), (\ref{44}) and (\ref{45}) which in principle
are supposed to be solved for $f_1$ and $f_2$. However, due to the nonlinear
structure of this set of coupled differential equations, and because of complexity of solving such equations, in practice we will not pursue the solving problem here.

\section{Some examples of generalized black holes}

In this section, following the above propositions, we aim to study some
generalized black hole solutions by expressing their metrics as multiply-warped product metric of generalized Friedmann-Robertson-Walker spacetime $\bar{M} = K\times_{f_1} M_1\times_{f_2}M_2$. We show the reduction of Einstein equations $G_{AB} = -\bar{\Lambda}g_{AB}$ into $G_{\alpha\beta} = -\Lambda_1g_{\alpha\beta}$ and $G_{ij} = -\Lambda_2 h_{ij}$, having support in $\bar{M}$, $M_1$ and $M_2$, respectively. In doing so, we consider some examples of well known black hole solutions and their possible generalizations such as {\it ``n-dimensional Schwarzschild black hole''}, {\it ``n-dimensional Reissner-Nordstrom black hole''} having
higher dimensional generalization, {\it ``Generalized (2+1)-dimensional Banados-Teitelboim-Zanelli (BTZ) black hole''} and {\it ``Generalized (2+1)-dimensional de Sitter black hole''} having Noether symmetry generalization.

\subsection{\bf n-dimensional Schwarzschild black hole}

An $n$-dimensional Schwarzschild black hole has the following metric \cite{Kono}
\begin{align}\label{SC}
&ds^2=-f(r)dt^2 + f(r)^{-1}dr^2 + r^2d\Omega_{n-2}^2~,
\end{align}
where $f(r)=(1-\frac{m}{r^{n-3}})$, $d\Omega^2_{n-2}=
\frac{(2\pi)^{(n-1)/2}}{\Gamma((n-1)/2)} ,\Gamma(1/2) = \sqrt{\pi}
,\Gamma(z + 1) = z\Gamma(z)$ and $m$ is a geometric mass indicating for the horizon radius.
Now, assuming $f(r)<0$, we define a new coordinate $\mu$ inside the black hole horizon, with the radius $r_H$ obtained by solving the equation $f(r)=(1-\frac{m}{r^{n-3}})=0$,
as
\begin{align}\label{ds}
&d\mu=\sqrt{-f(r)^{-1}}dr~,~~~0\le r \le r_H,
\end{align}
which yields
\begin{align}
\mu=\int_{0}^r \sqrt{-f(r)^{-1}}dr=F(r).
\end{align}
Using $r=F^{-1}(\mu)$ and (\ref{ds}) in (\ref{SC}) we have
\begin{align}\label{ds^2}
&ds^2=-d\mu^2 + \left(\frac{m}{(F^{-1}(\mu))^{n-3}}-1\right)dt^2 +
(F^{-1}(\mu))^2
d\Omega_{n-2}^2~.
\end{align}
Comparing (\ref{ds^2}) with an $n$-dimensional multiply-warped metric of
generalized Friedmann-Robertson-Walker spacetime $\bar{M}=K\times_{f_1}M_1\times_{f_2} M_2$ as
\begin{align}\label{MWM}
&ds^2=-d\mu^2+f^{2}_1(\mu)dt^2+f^{2}_2(\mu)d\Omega_{n-2}^2~,
\end{align}
we obtain
\begin{align}
&f_1(\mu)=\sqrt{\frac{m}{(F^{-1}(\mu))^{n-3}}-1}~,\\
&f_2(\mu)=F^{-1}(\mu).
\end{align}
The existence of above functions $f_1(\mu)$ and $f_2(\mu)$ guarantees the reduction of Einstein equations $G_{AB} = -\bar{\Lambda}g_{AB}$ into
$G_{\alpha\beta} = -\Lambda_1g_{\alpha\beta}$ and $G_{ij} = -\Lambda_2 h_{ij}$, where $\bar{\Lambda}$, $\Lambda_1$, and $\Lambda_2$ are considered as cosmological constants subject to the set of coupled differential equations (\ref{BarLambda}), (\ref{44}) and (\ref{45}) with the replacement of $t$ by $\mu$.

\subsection{\bf n-dimensional Reissner-Nordstrom black hole}

In dimensions $n\geq4$, a Reissner-N¨ordstrom black hole has the metric
\begin{align}\label{RN}
ds^2=-f(r)dt^2+f(r)^{-1}dr^2+r^2d\Omega_{n-2}^2,
\end{align}
where $f(r)=(1-\frac{m}{r^{n-3}}+\frac{q}{r^{2(n-3)}})$, $m$ and $q$ are
the geometric mass and charge of the black hole, respectively, and
$\Omega_{n-2}=\frac{2\pi}{\Gamma(\frac{n-1}{2})}$.
Similar to Schwarzschild black hole we may define a new coordinate
\begin{align}
&d\mu=\sqrt{-f(r)^{-1}}dr~,~~~r_-\le r \le r_+,
\end{align}
where $r_-$ and $r_+$ are the inner and outer horizons of the charged black
hole obtained by solving the equation $f(r)=(1-\frac{m}{r^{n-3}}+\frac{q}{r^{2(n-3)}})=0$.
Then, the metric (\ref{RN}) is written as an $n$-dimensional multiply-warped metric of generalized Friedmann-Robertson-Walker spacetime
\begin{align}
ds^2=-d\mu^2+f^{2}_1(\mu)dt^2+f^{2}_2(\mu)d\Omega_{n-2}^2~,
\end{align}
where
\begin{align}
&f_1(\mu)=\sqrt{\frac{m}{(F^{-1}(\mu))^{n-3}}-\frac{q}{(F^{-1}(\mu))^{2n-6}}-1}~,\\
&f_2(\mu)=F^{-1}(\mu)~,
\end{align}
with
\begin{align}
\mu=\int_{r_-}^r \sqrt{-f(r)^{-1}}dr=F(r).
\end{align}
The existence of above functions $f_1(\mu)$ and $f_2(\mu)$ guarantees the reduction of Einstein equations $G_{AB} = -\bar{\Lambda}g_{AB}$ into
$G_{\alpha\beta} = -\Lambda_1g_{\alpha\beta}$ and $G_{ij} = -\Lambda_2 h_{ij}$, where $\bar{\Lambda}$, $\Lambda_1$, and $\Lambda_2$ are considered as cosmological constants subject to the set of coupled differential equations (\ref{BarLambda}), (\ref{44}) and (\ref{45}) with the replacement of $t$ by $\mu$.

\subsection{ \bf Generalized $(2+1)$-dimensional  Banados-Teitelboim-Zanelli (BTZ) black hole}

Ba$\tilde{n}$ados, Teitelboim and Zanelli have shown that $(2+1)$-dimensional gravity with a negative cosmological constant has a black hole solution, so called BTZ black hole \cite{BTZ}. In Ref.\cite{Hong}, it has been shown that the BTZ black hole can be written as a multiply-warped metric of generalized Friedmann-Robertson-Walker spacetime. Recently, the BTZ black hole has been generalized using the Noether symmetry \cite{DAR}. In this section, we aim to find the multiply-warped product metric form of this generalized
BTZ black hole. The line element of the generalized (2+1)-dimensional BTZ black hole solution having Noether symmetry is expressed as \cite{DAR} \\
\begin{align}\label{59}
ds^2 = -f^2(r) dt^2 + f^{-2}(r) dr^2 + r^2 [ g(r)dt + d\phi]^2,
\end{align}
$$
f^2(r) = -m+\frac{r^{2}}{l^{2}}+\frac{J^2}{4r^2}+\frac{2Q}{r},
$$
$$
g(r)=-\frac{J}{2r^2},
$$
where $m$, $J$ and $Q$ are the mass, angular momentum and Noether charge
of the black hole.
Now, we define a new coordinate $\mu$ as
\begin{align}\label{60}
&d\mu=\sqrt{-f(r)^{-2}}dr~,~~~r_-\le r\le r_+,
\end{align}
where $f(r)^2<0$ and $r_-$, $r_+$ are respectively the radius of inner and outer horizons of the generalized $(2+1)$-dimensional BTZ black hole, obtained
as the solutions of $f^2(r)=(-m+\frac{r^{2}}{l^{2}}+\frac{J^2}{4r^2}+\frac{2Q}{r})=0$ \cite{DAR}. Eq.(\ref{60}) can be integrated to yield
\begin{align}
\mu=\int_{r_-}^r \sqrt{-f(r)^{-2}}dr=F(r).
\end{align}
Considering the new coordinate $\mu$, we can express the metric of generalized $(2+1)$-dimensional BTZ black hole as the metric of multiply-warped product form
\begin{align}\label{69}
&ds^2=-d\mu^2+f^{2}_1(\mu)dt^2+f^{2}_2(\mu)[g(r)dt + d\phi]^2~.
\end{align}
To cast this form into the form of multiply-warped product metric (\ref{MWM}) in a {\it comoving coordinates}, one can replace $[g(r)dt + d\phi]$ by $d\phi$ \cite{Hong} in (\ref{69}) to obtain the modified $f_1(\mu)$ and $f_2(\mu)$ as
\begin{align}
&f_1(\mu)= \left( -m+\frac{(F^{-1}(\mu))^{2}}{l^{2}}+\frac{J^2}{4(F^{-1}(\mu))^2}+\frac{2Q}{(F^{-1}(\mu))} \right)^{1/2},\\
&f_2(\mu)=F^{-1}(\mu).
\end{align}
The existence of above functions $f_1(\mu)$ and $f_2(\mu)$ guarantees the reduction of Einstein equations $G_{AB} = -\bar{\Lambda}g_{AB}$ into
$G_{\alpha\beta} = -\Lambda_1g_{\alpha\beta}$ and $G_{ij} = -\Lambda_2 h_{ij}$, where $\bar{\Lambda}$, $\Lambda_1$, and $\Lambda_2$ are considered as cosmological constants subject to the set of coupled differential equations (\ref{BarLambda}), (\ref{44}) and (\ref{45}) with the replacement of $t$ by $\mu$.

\subsection{\bf Generalized (2+1)-dimensional de Sitter black hole}

In this section, we find the multiply-warped product metric associated with the de Sitter three-metric (\ref{59}), outside the horizon with the generalized lapse function
\begin{align}\label{N^2}
f^2(r) = m-\frac{r^{2}}{l^{2}}-\frac{J^2}{4r^2}-\frac{2Q}{r}.
\end{align}
We define a new coordinate $\mu$ which is the same as generalized BTZ case
except that it is defined outside the horizon
\begin{align}
&d\mu=\sqrt{-f(r)^{-2}}dr~,~~~r> r_+,
\end{align}
which results in
\begin{align}
\mu=\int_{r_+}^r \sqrt{-f(r)^{-2}}dr=F(r),
\end{align}
where $f(r)^2>0$. Using $r=F^{-1}(\mu)$ in the de Sitter three-metric (\ref{59})
with the generalized lapse function (\ref{N^2}) and {\it comoving coordinates} ($g(r)dt + d\phi\rightarrow d\phi$) we have
\begin{align}\label{align}
&ds^2=-d\mu^2 +\left(m-\frac{(F^{-1}(\mu))^{2}}{l^2}-\frac{J^2}{4(F^{-1}(\mu))^{2}}-\frac{2Q}{F^{-1}(\mu)}\right)dt^2 +(F^{-1}(\mu))^{2} d\phi^2.
\end{align}
Comparing (\ref{align}) with the $(2+1)$-dimensional multiply-warped product
metric
\begin{align}
&ds^2=-d\mu^2+f^{2}_1(\mu)dt^2+f^{2}_2(\mu)d\phi^2~,
\end{align}
we obtain
\begin{align}
&f_1(\mu)= \left(m-\frac{(F^{-1}(\mu))^{2}}{l^2}-\frac{J^2}{4(F^{-1}(\mu))^{2}}-\frac{2Q}{F^{-1}(\mu)}\right)^\frac{1}{2},\\
&f_2(\mu)=F^{-1}(\mu).
\end{align}
The existence of above functions $f_1(\mu)$ and $f_2(\mu)$ guarantees the reduction of Einstein equations $G_{AB} = -\bar{\Lambda}g_{AB}$ into
$G_{\alpha\beta} = -\Lambda_1g_{\alpha\beta}$ and $G_{ij} = -\Lambda_2 h_{ij}$, where $\bar{\Lambda}$, $\Lambda_1$, and $\Lambda_2$ are considered as cosmological constants subject to the set of coupled differential equations (\ref{BarLambda}), (\ref{44}) and (\ref{45}) with the replacement of $t$ by $\mu$.



\end{document}